# Online Event Integration with StoryPivot

Anja Gruenheid[*]
[*] Systems Group
Dep. of Computer Science
ETH Zurich
anja.gruenheid@inf.ethz.ch

Donald Kossmann[*][†]
[†] Microsoft Research
donaldk@inf.ethz.ch

Divesh Srivastava
AT&T-Labs Research
divesh@research.att.com

## ABSTRACT

Modern data integration systems need to process large amounts of data from a variety of data sources and with real-time integration constraints. They are not only employed in enterprises for managing internal data but are also used for a variety of web services that use techniques such as entity resolution or data cleaning in live systems. In this work, we discuss a new generation of data integration systems that operate on (un-)structured data in an online setting, i.e., systems which process continuously modified datasets upon which the integration task is based. We use as an example of such a system an online event integration system called STORYPIVOT. It observes events extracted from news articles in data sources such as the 'Guardian' or the 'Washington Post' which are integrated to show users the evolution of real-world stories over time. The design decisions for STORYPIVOT are influenced by the trade-off between maintaining high quality integration results while at the same time building a system that processes and integrates events in near real-time. We evaluate our design decisions with experiments on two real-world datasets and generalize our findings to other data integration tasks that have a similar system setup.

## 1. INTRODUCTION

Data integration is traditionally seen as the task of integrating multiple types of data across various data sources often within enterprises. In recent years, research has shifted its focus from the traditional data integration setup, i.e., static datasets and batch processing, to continuous, incremental (and therefore online) data integration techniques [18]. That is, data integration systems are now expected to be online systems that propagate updates to the original datasets in near real-time into the integrated solution. As a result, these systems are required to deliver high quality integration in a timely and scalable manner without blocking user requests to the system. For example, engines that deploy entity resolution frameworks are the Facebook Entities Graph [19] and the Google Knowledge Graph [21]. They simultaneously extend their entity base by integrating new crowdsourced or crawled data about entities as more information becomes available while providing access to these entities through their public interfaces. Expanding the first generation data integration systems to online scenarios is an initial step in moving towards a next generation data integration system that is designed for efficient online integration. However, we argue that we further need to expand the scope of such systems beyond structured, well-defined data such as entities. Specifically, a large percentage of online data is available as unstructured data. Therefore, the next generation of online data integration systems should be able to support unstructured data containing structured concepts such as entities.

**Problem.** This paper focuses on a typical application for this type of novel data integration system which is an event processing and integration system. The goal of that system, STORYPIVOT, is to integrate real-world events into so-called *stories* which describe the continuous relationship between events over time. Take as an example the refugee crisis in Europe. Here, news coverage from different data sources connects news articles about the conditions in a Greek refugee camp to interviews of people escaping Syria to information about German immigration politics. The challenge here is to design a system that can process and correlate this kind of events over time and over multiple data sources (e.g., newspapers, blogs etc.). Furthermore, this integration needs to be achieved in a streaming setup because events are continuously reported. There exists a trade-off between the quality of the integrated result and system performance, i.e., how fast data can be integrated, which we examine in our work. We will furthermore show that the developed techniques can benefit a larger set of online data integration applications on structured and unstructured data.

**Approach.** STORYPIVOT leverages three observations to address the problem of online data integration. First, data within a data source commonly conforms to a source-specific standard and typically has high logical consistency within that source. For example in newspapers, the same real-world event is not covered by multiple news articles unless these articles describe a different angle or additional information becomes available. Thus, there exists an implicit notion of continuity within a data source which can be leveraged to provide high quality integration results over time. Second, if we leverage this source-specific observation, we can parallelize the data integration process. Specifically, STORYPIVOT first integrates data within a data source which is independent of the data integration process in any other data source. The resulting per-source stories are then integrated across data sources to provide a holistic view. Third, in addition to inter-source parallelization, we leverage intra-source parallelism to reduce the latency of data integration. This paper describes the design decisions that were made when creating STORYPIVOT and will highlight useful techniques for enabling parallelized, online data integration systems.

**Contributions.** To the best of our knowledge, this is the first work that explicitly describes how to construct an online, scalable data integration system from scratch and that discusses which design decisions influence system performance and result quality. With our work, we make the following conceptual contributions.

**Online Data Integration.** We introduce the notion of continuous data integration systems that are suitable for linking large amounts of data in a scalable manner.

**StoryPivot Design.** To show the design considerations for an online data integration system, we present a novel event process-ing and integration system called STORYPIVOT that provides integrated data for the user through scalable, near-real time data processing mechanisms.

**Generalization.** We generalize our findings to a wider range of systems that focus on linking data over time.

The remainder of this paper is organized as follows. First, we position our work in the context of existing work in the area of data integration in Section 2 and give an overview of the design decisions incorporated in STORYPIVOT. We then explain the terminology used throughout this paper in Section 3. In Section 4 we discuss the impact of our design decisions on the integration quality of our system while Section 5 focuses on the system performance. Our design decisions are thoroughly evaluated empirically in Section 6. Finally, we generalize our findings in Section 7 and examine how previous research relates to our findings in Section 8.

## 2. PROBLEM STATEMENT

Traditional data integration typically focused on offline data processing, integrating snapshot data, e.g., in ETL processes. In that context, integration commonly consisted of three steps: 1) schema alignment to understand the vocabulary of the integration process, 2) record linkage to determine common entities, and finally 3) data fusion to ensure consistency. STORYPIVOT can be described as an online linkage system that allows for continuous data integration over time. In contrast to traditional integration techniques, it produces integration results incrementally while using techniques inspired by blocking to improve system performance. Schema alignment in STORYPIVOT is done as part of the data extraction phase, i.e., we preprocess textual data and annotate it with the same vocabulary across data sources. This allows the system to operate on structured data internally while enabling unstructured data as a potential input source. In the second step, record linkage, our system correlates data within and across data sources. The output of STORYPIVOT are so-called stories which describe the evolution of real-world events over time. Events are linked but not fused when constructing stories to ensure that source biases are preserved and available for subsequent analyses. However, fusion or summarization of stories is a natural extension of our work.

There exist two areas of research that are closely related to *online event integration* which is the core mechanism applied in STORYPIVOT. First, *incremental data integration* is a research area that focuses on adapting integration algorithms that run on large datasets to online data. Second, *data integration with evolving data* has emerged as a research area recently discussing continuous changes to integrated datasets. We outline both of these research areas next.

**Incremental Data Integration.** Online data integration has been an area of interest over the last years as research has increasingly focused on modifying traditional batch algorithms [9, 23] to suit online data integration settings for record linkage [10, 22, 40] or data fusion [29]. These modifications limit the batch execution to only those portions of the data integration solution that have been (indirectly) changed or compute the answer incrementally until (near-)certainty is reached. They aim to improve performance while maintaining good integration quality. However, their goal is to modify the content of entities or objects rather than document their evolution over time.

**Data Integration with Evolving Data.** Evolution is a concept that is most commonly studied for entities such as people, organizations, etc. As an example take temporal clustering which is a technique that has been recently employed for record linkage [13, 28]. For example, a professor may take on an appointment at a different university. As she is still the same person, a record linkage system should be able to correlate multiple records that contain this kind of varying but evolving information. Furthermore, this problem has been extended allowing entities to evolve over time after which they are clustered to form entity-dependent stories [6] and has also been studied for evolving correlation between textual data describing entities [12]. This core idea is analogous to the idea of the evolution of events over time. However, the problem presented in our work is multi-faceted while previous research has mostly focused on entities only. Specifically, for a story to evolve, it is not only the entities but also the topics, sentiments etc. that may evolve.

Table 1: Example of events that are processed by STORYPIVOT.

| ID | $t_i$ | Entities | Topics | Title |
|---|---|---|---|---|
| $r_1^1$ | Aug $12^{th}$ | Kos, Refugees | Politics, War | Migrants locked in stadium on Kos |
| $r_2^1$ | Aug $12^{th}$ | Kos, Refugees | Politics | Syrian refugees arrive on Kos |
| $r_3^1$ | Aug $12^{th}$ | Spain | People, Politics | Bullfighting returns to San Sebastian |
| $r_4^1$ | Aug $12^{th}$ | China | Disaster | Tianjin blast sets off earthquake |
| $r_5^1$ | Aug $13^{th}$ | China, Tianjin | Disaster | China blasts: hundreds injured |
| $r_6^1$ | Aug $13^{th}$ | Greece, Kos | Politics, War | Greece sends cruise ship |
| $r_7^1$ | Aug $13^{th}$ | Japan, Tianjin | Disaster | Tianjin explosions visible from space |
| $r_8^1$ | Aug $13^{th}$ | Italy | Crime, Politics | Mafia neighbours are bad for business |
| $r_9^1$ | Aug $14^{th}$ | Greece | War | Migrants: 'They said they'd give us papers' |
| $r_1^2$ | Aug $12^{th}$ | Isis | Politics, War | Islamic State vs Kurds: What's going on? |
| $r_2^2$ | Aug $12^{th}$ | Refugees, Turkey | People, War | Refugees in Turkey: 'Nothing for us here' |
| $r_3^2$ | Aug $13^{th}$ | Refugees, Greece | Politics, War | Chaos amid Greek registration attempt |
| $r_4^2$ | Aug $13^{th}$ | Kos, Refugees | People | What happens to migrants arriving on Kos? |
| $r_5^2$ | Aug $13^{th}$ | China, Tianjin | Disaster | Tianjin rocked by explosions |

The goal of STORYPIVOT is to establish a correlation between real-world events across time and across data sources, forming stories. To accurately represent what happens in the real world, the system first obtains a structured digital representation of different recorded events which we refer to as *snippets*. It then uses similarity computation and incremental clustering to correlate the snippets and to maintain them and their relationships continuously. These clusters thus reflect real-world stories. Computing stories from events is a novel concept introduced in STORYPIVOT, however, there exist several event processing systems that are designed for event detection and which are closely related to STORYPIVOT.

**Event Processing Systems.** These systems are applicable in a variety of domains, amongst others political science, finance, and news. Examples of existing systems are GDELT [11] and EventRegistry [26] (see Section 8 for additional examples). STORYPIVOT differs from GDELT as it has a broader scope than just political science and is designed to capture the evolution of events and not only the events themselves. Similar to EventRegistry, it examines news articles but instead of focusing on event identification, STORYPIVOT establishes the historical context of events. We discuss further applications of event processing systems in Section 8.

To understand the differences between STORYPIVOT and existing systems, take as an example Table 1 which shows snippets that correspond to events extracted between August 12th and 14th 2015 from the 'Guardian' (data source $s_1$) and 'BBC News' (data source $s_2$). Each snippet is annotated with a timestamp, associated entities, topics, and the title of the article. This information is obtained by taking the underlying unstructured news article and using an open-source NLP framework called NLTK [2] which annotates the article with structured data. We refer to each of the categories of the annotated data as a *dimension*. Then, a snippet $r_i^j$ denotes the i-th recorded snippet within the j-th data source. As an example, $r_2^1$ is the second snippet recorded in the 'Guardian' containing information on the arrival of Syrian refugees on a Greek island called Kos. We now outline the workflow of STORYPIVOT which shows how it integrates snippets and constructs their corresponding stories. This workflow is then discussed in detail throughout the remainder of this paper.

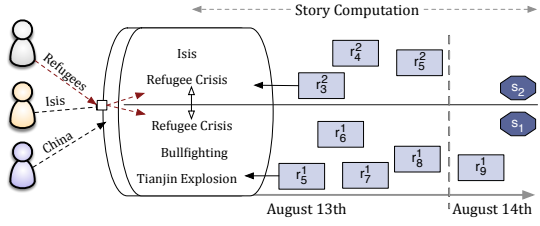

Figure 1: Workflow of STORYPIVOT.

STORYPIVOT **Workflow.** Snippets that are added to the system are processed in two steps. First, they are aggregated into clusters within the data source from which they originate. These clusters correspond to real-world stories as seen from the perspective of a single data source. In the second step, source-specific clusters are aligned and integrated across data sources. This allows the system to determine a global view of a story over time and also has the potential to fix data quality issues, for example if a real-world story is split into two clusters in the same source in the first step. An overview of STORYPIVOT is shown in Figure 1. Here, the snippets from August $12^{th}$ have already been processed and have been separated into three different clusters within data source $s_1$ and two different clusters within $s_2$. Furthermore, the system identified one aligned cluster representing a story about the refugee crisis across data sources. Newly arriving snippets throughout August $13^{th}$ and $14^{th}$ are managed through a continuously running backend process which applies the two-step integration policy for each snippet. These processes are transparent to the user who should be able to simultaneously access STORYPIVOT for information about stories.

## 3. TERMINOLOGY

In this section, we establish the terminology used throughout our work. Specifically, we show the mapping between real-world objects and their digital representations formally and also introduce the data structures that we use.

### 3.1 Definitions

As mentioned previously, the digital version of an event is a snippet which is a standardized excerpt of a news article that contains multiple dimensions, i.e., various types of content that describe the event. Going back to the news excerpts recorded in Table 1, $r_1^1$ describes the situation of Syrian refugees arriving in Kos as published by the 'Guardian' on August 12th 2015. This snippet further contains metadata that can be used for similarity computation such as a title ('Migrants locked in Stadium'), entities ('Kos', 'Syrian refugees'), or topics ('Politics', 'War').

**Sketches.** Let $R = \bigcup R_j$ be the set of snippets that the system observes across all data sources $s_j$. Computing the relationship between all snippets within a data source and across data sources is costly. Therefore, we use the concept of aggregation to efficiently reduce the number of comparison objects in the system within specified time windows. We refer to these aggregation objects as *temporal sketches* or *sketches* in short. A sketch $v_i^{j,t_k}$ represents a set of snippets $r_1^j \ldots r_l^j \in R$ which occur in the same data source $s_j$ during a time window $t_k$ and have similar semantic content. Each $v_i^{j,t_k}$ contains the same number of dimensions $d$ as any snippet $r_\ell^j \in R_j$ though their content may vary depending on the aggregation mechanism per dimension. STORYPIVOT as of now supports full aggregation (all dimensional data from the snippets is completely stored in the sketch) and top-k aggregation. Top-k sketches are summary sketches that do not contain every piece of information but only a selected subset such as the most important $k$ entities or the $k$ words that characterize the text of the snippets best. Our definition of a top-k sketch is loosely based on the notion of bottom-k sketches [14]. However, instead of applying a hash function to randomly select representative tokens for each sketch, we apply either cosine similarity with respect to the token's dimension or frequency counting to determine the top tokens. For example 'Refugees' may be a token that appears often in the 'entity' dimension of our running example while it is less common in the 'text' dimension. As a result, it should have more impact when mentioned in that dimension. The similarity between a sketch and a snippet can then be measured using (weighted) string similarity metrics or cosine similarity across these different dimensions. Details about the aggregation of snippets and its impact on system performance and result quality are further discussed in Section 4 and Section 5.

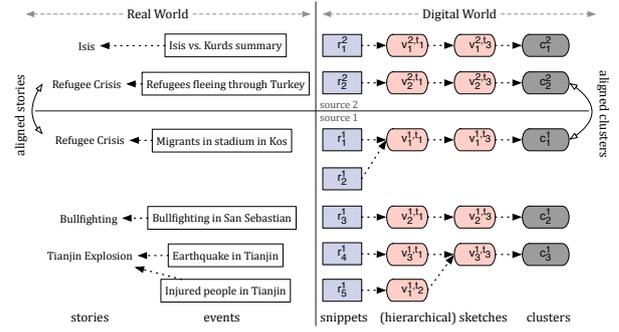

Figure 2: Mapping of real-world to digital-world terminology.

Sketches are the basic computational unit within STORYPIVOT. As mentioned above, they are computed for a time window $t_k$ which is a fixed period of time, for example a 24 hour interval which we will use throughout this paper in our examples. Within $t_k$, snippets that are semantically similar are integrated into one sketch as shown for the running example in Figure 3. Here, $r_1^1$ and $r_2^1$ are similar (they both contain information on the situation of Syrian refugees on Kos) and fall into the same time window, August $12^{th}$. They are therefore aggregated into the same sketch which is also visualized in Figure 2.

**Clusters.** If there exist two sketches that are sufficiently similar across time windows in data source $s_j$, we say that they belong to the same cluster $c_i^j$. In our running example, $v_1^{1,t_1}$ and $v_2^{1,t_2}$ as well as $v_3^{1,t_1}$ and $v_1^{1,t_2}$ respectively have similar content in their dimensions which results in them being added to $c_1^1$ ($c_3^1$). Clusters are the digital representation of stories while sketches represent one specific real-world event within a time window $t_k$. Furthermore, we allow sketch hierarchies, i.e., sketches can be combined if they are semantically similar within a time window $t_l$ and their associated time windows $t_{k_\ell}$ lie within $t_l$. The reason for generating hierarchies is system performance. As mentioned previously, comparing snippets (and sketches) exhaustively is time consuming. Reducing the number of sketches in the comparison space has immediate performance benefits although it may have negative impact on integration quality. We argue that the choice of aggregation level as well as hierarchy is dependent on the environment the system is used in.

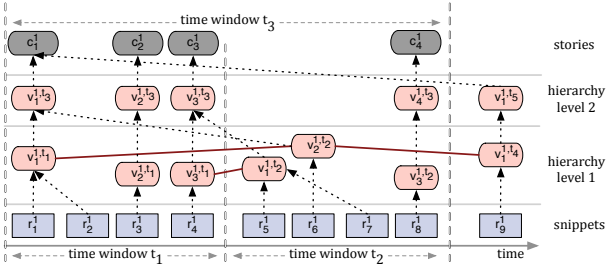

Figure 3: Relationship between snippets, temporal sketches, and clusters for snippets $r_1^1 - r_7^1$ of the running example (Table 1).

For example if communication across data sources is expensive, an intuitive solution is to use sketches that span a bigger time interval to exchange fewer messages. On the other hand, if the system is to use fine-grained sketches within a data source to depict the cluster evolution as accurately as possible, it is not possible to simply adjust the overall sketch size. If hierarchies are applied, the system can be tuned according to the system's workload which we exemplify for two different datasets in Section 6. Finally, an aligned cluster is the result of linking clusters $c_i^{s_j}$ across data sources and represents all knowledge that the system has about a specific real-world story.

## 3.2 Data Structures

To internally represent the relationship of sketches as well as the relationship of clusters across data sources, we leverage graph representations. Specifically, we model the intra-source relationship of sketches through a *sketch relationship graph* on top of which a clustering algorithm is applied to find the set of clusters $C_j$ that represent the stories within data source $s_j$. Inter-source relationships of clusters are stored in a *cluster relationship graph*. Applying a (potentially different) clustering algorithm on top of this graph, this graph contains aligned clusters, i.e., it enables us to find stories mentioned in multiple data sources. For the purpose of this paper, we use transitive closure as the clustering algorithm of choice for both types of clustering. However, it is part of our future work to determine whether different, temporally aware, clustering algorithms can further benefit the quality of STORYPIVOT's integration results.

**Sketch relationship graph.** To identify clusters within a data source, we use a sketch relationship graph $SRG_j = (V_j, E_j)$ where a sketch $v_i^{j,t_k}$ is a node in $V_j$ and an edge label $e(v_i^{j,t_k}, v_\ell^{j,t_l}) \in E_j$, with $e(v_i^{j,t_k}, v_\ell^{j,t_l}) \in (0, 1)$, is the similarity between two sketches $v_i^{j,t_k}$ and $v_\ell^{j,t_l}$ in data source $s_j$. If $e(v_i^{j,t_k}, v_\ell^{j,t_l}) = 1$ we say that $v_i^{j,t_k}$ and $v_\ell^{j,t_l}$ are identical. If $v_i^{j,t_k}$ and $v_\ell^{j,t_l}$ are semantically distinct, $e(v_i^{j,t_k}, v_\ell^{j,t_l}) = 0$ holds instead. For creating clusters within $s_j$, the system needs to identify those edges between $v_i^{j,t_k}$ and $v_\ell^{j,t_l}$ that indicate that they are similar. We say that there is sufficient evidence for $v_i^{j,t_k}$ and $v_\ell^{j,t_l}$ to belong to the same cluster if their similarity exceeds a threshold $\alpha_v$. This methodology is applied in a variety of graph clustering algorithms such as correlation clustering or cut clustering to reduce the clustering search space [10]. An example of a sketch relationship graph is the hierarchy level 1 layer of Figure 3. Here, an edge exists between $v_1^{1,t_1}$ and $v_2^{1,t_2}$ because both of these sketches contain information about entity 'Kos' as well as topics 'Politics' and 'War'. Additionally, $v_2^{1,t_2}$ has an edge to $v_1^{1,t_4}$ because both of these sketches overlap in the entity dimension and partially overlap in the topics dimension whereas $v_1^{1,t_1}$ only partially overlaps with $v_1^{1,t_4}$ in both dimensions. Analogously, there exists an edge between sketches $v_3^{1,t_1}$ and $v_1^{1,t_2}$. For the sake of readability, assume in the following that edges take a value of 1 or 0, i.e., there is sufficient evidence that sketches are or are not similar.

**Cluster relationship graph.** A cluster relationship graph is a graph $CRG = (C, H)$ where a node $c_i^j \in C$ represents a source-specific cluster and an edge in the edge set $H$ denotes the similarity between two clusters $c_i^j$ and $c_\ell^l$, $j \neq l$. In other words, a cluster relationship graph describes the relationship of clusters across data sources. Analogous to the sketch relationship graph, we say that clusters are similar across data sources if they exceed a similarity threshold $\alpha_c$. An example of an aligned cluster is shown in Figure 2 on the right-hand side where the two clusters representing the refugee storyline are connected across clusters.

## 4. INTEGRATION QUALITY

We next detail the two-tier architecture of STORYPIVOT that allows us to efficiently integrate snippets into existing clusters. We call these two tiers *cluster identification* and *cluster alignment*.

## 4.1 Cluster Identification

Cluster identification refers to the construction of clusters, i.e, the equivalent of real-world stories, within a data source. To find the cluster that a snippet may belong to, we follow a three-step process. First, we identify the sketches in the system that are similar to the extracted snippet. Second, we merge the snippet with the best match or generate a new sketch if STORYPIVOT did not find a matching sketch. Third, the (new) sketch is compared to other sketches across time windows to identify how similar it is to them. If sufficiently similar sketches are identified, we trigger a reclustering of the sketches.

### 4.1.1 Snippet Matching

As described previously, the content of a snippet can be divided into multiple dimensions. Dimensions, for example entities, topics, title etc., are the access points for snippet matching. Specifically, a snippet can only be similar to a sketch if it has overlap with the sketch in one or more dimensions. As an example, recall Table 1. After $r_1^1$ is integrated into STORYPIVOT, the system contains exactly one sketch $v_1^{1,t_1}$ with (a subset of) the dimension content of $r_1^1$. When $r_2^1$ is added to the system, it identifies $v_1^{1,t_1}$ as a candidate match because their dimensions are similar (the entities are the same and its topics are a subset of $r_1^1$).

**Implementation.** To be able to fetch all candidate sketches within the time window corresponding to the incoming snippet, the system deploys a multi-level hash-based data structure. It contains metadata per dimension, time window, and data source that have been observed in the system as shown in Figure 4. Indices are not collapsed for one straightforward reason: The distribution of tokens (words, entities, topics etc.) varies depending on the dimension. For example the distribution of words in the titles of these snippets is less constrained than either the topic or entity dimension. As a result, the similarity computation should differ across dimensions which can be achieved by recording dimensions separately. Furthermore, this allows us to parallelize access to the indexing structure: as retrieval usually accesses different parts of the indexing structure, these accesses can be executed in parallel. To reduce the similarity computation space, i.e., to avoid comparing all sketches that match for just one dimension, we use a bloom filter and a threshold for the minimum number of matching dimensions. This filtering mechanism is often helpful as dimensions may have low selectivity. An example is the topic dimension in our running example: there are only a limited number of assigned topics. Without the filter, this would cause a significant computational overhead otherwise.

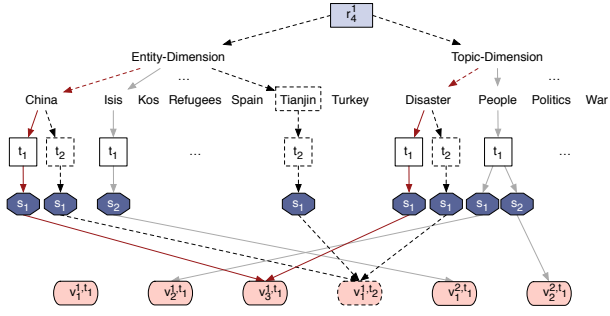

Figure 4: Partial index for dimensions 'entity' and 'topic' of the running example for snippets $r_1^1 - r_4^1$ and $r_1^2 - r_2^2$.

EXAMPLE 1. *To retrieve candidate matches, we traverse the index as follows. For each snippet dimension, we extract its tokens, for example {China, Tianjin} in dimension 'Entity' for snippet $r_4^1$. Next, we retrieve all sketches within the same time window that contain the same set of tokens in that dimension. In Figure 4, we show that for the 'Entity' dimension, the only match for data source $s_1$ and token 'China' is for timestamp $t_1$. Analogously, there exists no matching sketch in dimension 'Topic' for this snippet. Therefore, the system generates a new sketch, $v_1^{1,t_2}$, that contains $r_4^1$ and adjusts the indexing structure accordingly.*

### 4.1.2 Snippet-Sketch Integration

Whether the system decides to merge a snippet into an existing sketch is dependent on the similarity between the snippet and the sketch, i.e., whether it exceeds the similarity threshold $\alpha_v$. Let $D$ be the dimensions for which the system compares the snippet and the sketch. We say that STORYPIVOT merges snippet $r_i^j$ into sketch $v_\ell^{j,t_k}$ if a) the timestamp of $r_i^j$ is within time window $t_k$ and b) the normalized sum of the similarities across the dimensions is bigger than $\alpha_v$, i.e., if the following requirement holds:

$$\frac{\sum_{d \in D} sim_d(r_i^j, v_\ell^{j,t_k}) * w_d}{\sum_d w_d} \geq \alpha_v \quad (1)$$

Here, we allow different dimensions to be annotated with different weights $w_d$. The reason is that some dimensions have a stronger signal than others for integration purposes. Specifically, dimensions such as entities are effective means of identifying actors in a story where the combination of actors is often a strong signal for the continuity of that story. In comparison, the title of a snippet can be misleading because it is subject to the writer's creativity. As of now, the weights and the similarity threshold $\alpha_v$ used in STORYPIVOT are manually tested and based on a subset of our data for which we have identified a gold standard manually. Extending this to an adaptive or learned weighting scheme is a part of future research. Ideas in that context encompass per-source modeling of vocabulary and its relevance as well as source-specific weighting schemes.

### 4.1.3 Cluster Construction

In contrast to topic-focused event processing systems such as GDELT [11], the idea behind STORYPIVOT is that stories may evolve over time, i.e., it is possible that not all parts (sketches) of the story contain the same dimensional data. Instead, the evolving nature of a story is recorded through the change in the semantic overlap of the dimensions. That is, sketches at time window $t_k$ and $t_l$ can belong to the same cluster even if their metadata are distinct but $t_k \ll t_l$ holds and there exists (at least) a sketch in $t_\ell$ that is sufficiently connected to the sketches in $t_k$ and $t_l$. As a result, to construct clusters, the system first needs to identify candidate sketches in other time windows that may be related to the current sketch. Afterwards, it computes their similarity and adjusts the clustering solution.

**Implementation.** The identification of candidate sketches that are similar to the new or modified sketch is analogous to the initial snippet matching step. We then compute the pair-wise similarities through a slightly modified version of Equation (1), calculating the similarity of two sketches instead of the snippet-sketch similarity. These similarities are then added to the sketch relationship graph. Afterwards, incremental clustering mechanisms as described in Section 2 are triggered on top of the part of the sketch relationship graph that has been modified. In this phase, one or multiple clusters may have changed and are thus marked. For each of these marked clusters, the cluster alignment phase is then triggered to readjust the inter-source cluster relationships.

EXAMPLE 2. *In Figure 4, the system determined that $r_4^1$ is assigned to a new sketch $v_1^{1,t_2}$. However, there exists another sketch, $v_3^{1,t_1}$ that has overlap with $v_1^{1,t_2}$ due to its similarity in the entity ('China') and topics dimension ('Disaster'). As a result, we decide in the clustering phase that these two sketches are assigned to the same cluster, $c_3^1$, which represents the 'Tianjin Explosion' storyline within data source $s_1$.*

### 4.2 Cluster Alignment

Aligning clusters across data sources benefits users in two ways. First, if clusters within some of the data sources are incomplete, they can be enriched when presented to the user as stories. Second, clusters might differ depending on the data source. For example a newspaper that traditional favors democratic politicians will assess a republican debate differently than a republican newspaper. Adding the information of multiple data sources to the output, i.e., presenting data from different data sources to the user, therefore has the potential to improve the knowledge gain of the interested user.

The computational effort spent on the alignment of clusters across data sources is correlated to the number of data sources that the system uses as input as well as the number of objects that need to be compared per data source. To reduce the time spent on aligning clusters, we introduced the notion of sketch hierarchies in Section 3. The idea of sketch hierarchies is to abstract the clusters that have been found but to maintain their temporal character. That is, if cluster summaries are used instead, the evolution over time may not be captured accurately which would thus lead to erroneous alignments. Obviously, any aggregation implies that (part of) the data is omitted. One of the questions that we address in our evaluation (see Section 6.3) is how to find a good balance between aggregation and computational effort.

**Implementation.** Cluster alignment is triggered after the cluster identification phase has been completed. At this point, the system knows which clusters have (not) been modified. For each of the top-level sketches of each of the modified clusters, the system then retrieves candidate sketches that match their metadata on the same hierarchy level. For this step, the same index as shown in Figure 4 can be used as it already splits the dimensional vocabulary by data source. For all candidate sketches, we then (re-)compute the similarity between the clusters. Specifically, the similarity is measured as the (weighted) average of all top-level aggregate sketches. If the similarity exceeds a certain threshold, we say that these clusters likely represent the same real-world story. This information is then stored in the $CRG$.

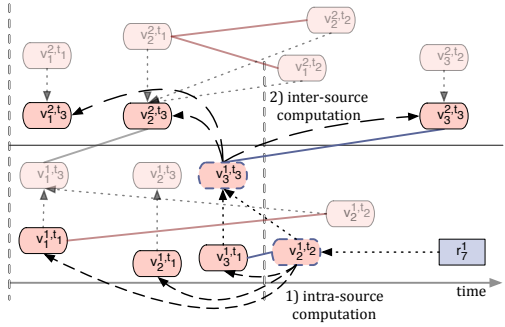

Figure 5: Intra- and inter-source computation in STORYPIVOT.

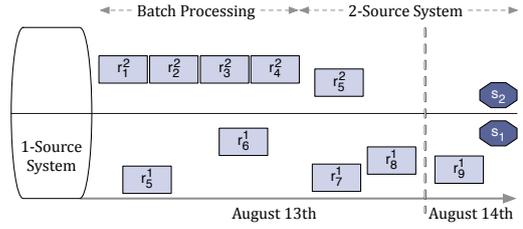

Figure 6: Online integration of data source $s_2$ into STORYPIVOT.

## 5. SYSTEM PERFORMANCE

The core concern when building a large scale online system is to provide high quality and up-to-date data items at any point in time. There is an inherent trade-off between data quality and system performance as more (detailed) data usually leads to better integration results but increases computation time. In the previous sections, we have discussed how we can aggregate data to minimize the number of comparisons between sketches (and summary sketches) that the system has to make. At the same time, our aggregation mechanisms emphasize a source-centric system structure. We further utilize this structure to address the two system challenges that we have described earlier, system latency and scalability. We next discuss how we can a) parallelize parts of our execution pipeline without loss of quality which decreases latency because of faster end-to-end processing times and b) how we can add and remove data sources from the system, thus addressing system scalability.

### 5.1 Pipeline Parallelization

In Section 4, we showed how separating data sources benefits the quality of integration results as the semantic cohesion within a data source is commonly higher than across data sources. Furthermore, we showed how we can improve system performance by aggregating snippets into sketches and finally clusters to decrease system load. Hierarchical compression is not the only measure that STORYPIVOT takes to improve system latency, i.e., the time it takes to fully integrate a snippet. Leveraging the source-centric data processing model, we discuss two types of parallelization that STORYPIVOT applies to improve processing time. First, we introduce the notion of *intra-source* parallelism which is to parallelize the identification of clusters within a data source. Second, *inter-source* parallelism is the task of parallelizing the computation of aligned clusters across data sources. The computation models for both types of parallelization are shown in Figure 5 and are described in detail next.

**Implementation of Intra-Source Parallelism.** Recall that in order to evaluate the similarity of a modified sketch with other sketches within a data source, the system compares it to all other candidate sketches that are *not* in the same time window. Similarity computation is inherently parallelizable as the score between two sketches is dependent only on the content of the sketches. Thus, STORYPIVOT computes these similarities as follows. For each time window $t_l$ that is different than the time window of the modified sketch $v_i^{j,t_k}$, the system assigns a thread from a pool of available threads that finds those sketches that have an overlap with $v_i^{j,t_k}$. Let $m$ be the total number of time windows currently stored in STORYPIVOT. We do not need to parallelize the execution for each of these time windows but only for the last $m'$ time windows. Recall that we assume an evolution of events, and thus sketches, over time. To model this evolution, it is not necessary to compare $v_i^{j,t_k}$ with all other sketches $v_\ell^{j,t_l}$. Instead, for a cluster to evolve, we only need to show that there is an overlap of $v_i^{j,t_k}$'s content with some other sketch $v_\ell^{j,t_l}$ if $v_\ell^{j,t_l}$ is in any of the last $\{t_{k-m'+1}, \ldots, t_k\}$ time windows. The number of parallel threads is thus reduced to $m'$ where commonly $m' \ll m$ holds. Afterwards, we collect the results of these similarity computations and synchronize the threads before initiating the next algorithm phase, cluster construction. Synchronization at this point prevents merging the same clusters multiple times. Specifically, it is likely that a new sketch is similar to several sketches that belong to the same cluster if it is similar to one of them. Thus, synchronization avoids triggering the same incremental clustering step multiple times. Depending on the cluster assignment, sketches and snippets are shuffled and hierarchical sketches recomputed. Afterwards, cluster alignment is initiated.

**Implementation of Inter-Source Parallelism.** Similarly to comparing sketches across time windows, their comparison across data sources is inherently parallelizable. That is, the pair-wise similarity of two sketches is not influenced by anything other than their semantic and syntactic relationships. Thus, it is possible to parallelize this kind of similarity computation without decreasing integration quality. In its framework, STORYPIVOT manages a second thread-pool for inter-source parallelization. If a cluster is modified and the alignment phase has been triggered, the system uses one thread per other data source out of that pool to calculate the similarity of the cluster's sketches with candidate sketches in the respective data source. Once these sketches have been found, their corresponding clusters are derived which is also an operation that can be executed in parallel. Finally, the system recalculates the similarity of the current cluster with each of these clusters and joins the threads to finish the execution.

### 5.2 Scaling StoryPivot

In addition to the parallelization of computational efforts within STORYPIVOT, its source-centric design enables us to easily add and remove data sources from the system. Take as an example the execution scenario shown in Figure 6 . Here, snippets $r_1^1$ - $r_4^1$ of data source $s_1$ have been inserted into the system which at that point contains a set of clusters constructed from these snippets. The integration of data source $s_2$ is now done in parallel to the (streaming) snippet integration of $s_1$: The system batch processes all snippets until it is up-to-date and then proceeds to integrate snippets in a streaming manner. The system load will obviously increase while STORYPIVOT tries to catch up the new data source to the current point in time. However, other data sources that are connected to the system are only minimally impacted by the additional load due to the source-centric design pattern which effectively separates the data sources. As we use separate thread pools for each data source, the computation is shifted to different cores in the machine which minimizes the disruption of the online system.

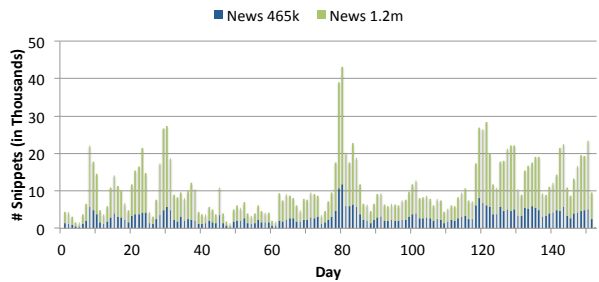

Figure 7: Distribution of snippets in the *News 465k* and *News 1.2m* and datasets (stacked).

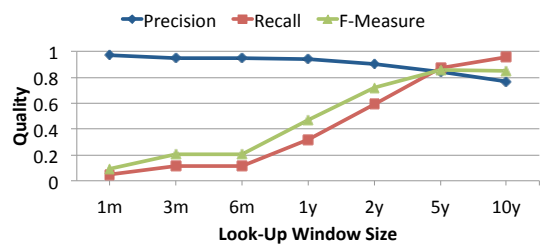

Figure 8: Quality with varying window sizes on the *Patents* dataset.

## 6. EXPERIMENTAL EVALUATION

In this section, we evaluate STORYPIVOT in terms of the trade-off between its end-to-end processing time and result quality. We first describe the setup that we use throughout our experimental evaluation. Afterwards, we examine the quality of our source-centric integration technique. Finally, we discuss the performance of STORYPIVOT and competing event processing techniques.

### 6.1 Setup

This section describes the two different datasets that were used for this evaluation as well as implementation details of STORYPIVOT and the experimental environment.

#### 6.1.1 Datasets

For this evaluation, we leverage two real datasets, *News* and *Patents*, in multiple variations as follows:

**News.** The *News* dataset has been collected over 151 days (1st October 2014 - 28th February 2015). It is extracted through the open-source backend of EventRegistry [26] and provides news articles extracted from various data sources. Amongst these are well-known newspapers such as 'The Guardian', and 'BBC', online news portals such as 'Yahoo News', and gossip data sources such as 'PerezHilton.com'. We extract one or multiple snippets per news article by using the NLTK [2] annotation tool. Each of the extracted snippets has five dimensions, a *title*, associated *people*, associated *organizations*, associated *locations*, and the original *text*. From the original dataset, we use three variations in this evaluation:

1. *News 465k* contains 465,321 snippets extracted from 30 different data sources.
2. *News 1.2m* contains 1,168,233 snippets extracted from 50 different data sources.
3. *Obama* contains 38,695 snippets that all have the term 'Obama' in their people dimension. This dataset is a subset of the *News 465k* dataset.

The distribution of snippets per day for the first two datasets is shown in Figure 7. We observe several patterns in these datasets: for example Saturdays and Sundays usually have a smaller number of observed events, there exist peaks that are caused by important events (right before Christmas (around day 80) significantly more news articles are published), and there are fewer snippets extracted in the mid October - November period than in the beginning of 2015. For the *Obama* dataset, we observe that the maximum number of processed snippets per day is 1227, the minimum 4, and on average 258 snippets are processed each day.

**Patents.** This dataset has been used to evaluate temporal data integration methods in prior research [28]. As such, it provides us with the means to evaluate our approach in a slightly different setting but where a ground truth is provided to determine the quality of our stories. A story here corresponds to the evolution of a person as recorded by her filed *name* and *address* for each invention. These may change over time. The dataset contains 1925 records with an average of 5.35 records per story in the ground truth.

#### 6.1.2 Implementation

We implemented STORYPIVOT (SP) as a single-server parallelized system as described previously. The implementation is done in Java and all experiments are run on the same server setup, a Linux machine with two Intel Xeon L5520 processors and 24GB RAM.

#### 6.1.3 Execution

Our streaming system mimics real-world streaming services. That is, each dataset is evaluated by imitating streaming behavior such that the execution is equivalent to the original evaluation of the dataset but at a faster speed. Specifically, for the *News* datasets, we obtain our results by simulating one day of the dataset within one execution hour. For the *Obama* dataset, we execute one week in one hour as the dataset contains a smaller number of snippets. Finally, we process ten years in a minute in the execution of the *Patents* dataset as it is comparatively small.

#### 6.1.4 Reference Systems

To compare the execution time of STORYPIVOT, we implemented a round robin scheme that is a naive alternative to the per-source processing model. In it, every snippet is processed using the same number of threads as used in STORYPIVOT in a round robin manner. We refer to this processing model in the following as ROUND. Second, we compared our implementation to a sequential processing model (SEQU) using the same story computation algorithms as the parallelized approaches. To compare the result quality of our parallelized approach, we determine the similarity between its result and the event clustering obtained by EventRegistry [26] and discuss our findings in Section 6.2.3.

### 6.2 Integration Quality

In this part of our evaluation, we focus on the integration quality of our system. We split our experiments into three parts. First, we examine the *Patents* dataset to determine the accuracy of our stories when compared to previously established temporal data integration techniques. Second, we compare the output of our system with EventRegistry which applies a domain-based online clustering algorithm [4] by manually comparing our stories with their extracted events. Finally, we compare our system to a sequentially executed baseline that applies the same merging policy as STORYPIVOT but does not differentiate between data sources.

#### 6.2.1 Patents Experiments

To the best of our knowledge, the *News* datasets do not have any associated ground truth which would allow us to evaluate the quality of our system. However, temporal entity resolution is a closely related topic for which several standardized datasets have been introduced over the last few years. The *Patents* dataset is one of them. It contains data extracted from the European Patent Registry where records are timestamped with the date the patent is filed and they also contain some (limited) information about who filed the patent. Thus, we can use this data to establish the patent filing story of a person. Furthermore, we use blocking as applied in the original temporal record linkage paper [28]. Specifically, the dataset is annotated such that persons starting with the same letter are processed in the same block. STORYPIVOT is a highly parallelized, sketch-based data integration system. To evaluate its quality, we can vary several parameters such as the degree of parallelization or aggregation. The top f-measure that we observe in our experiments is comparable to the reported f-measure for optimized temporal record linkage techniques as shown in [28] which indicates that there is no loss of quality because of our parallelized execution framework. We discuss our observations for different experimental setups on this dataset next.

**Intra-Source Parallelization.** A larger window size, i.e., the span we go back in time to find sketches that may match the current story or sketch, means more computational effort which we verify in our performance evaluation (see Section 6.3 for details). However, the nature of the *Patents* dataset is that patents are filed on a yearly rather than daily or even weekly basis as we would assume for data such as contained in the *News* datasets. Thus, we obtain the top f-measure of 0.86 when the window size is five years. The trade-off for different window sizes in terms of precision, recall, and f-measure is shown in Figure 8. If the window size chosen is too small, stories cannot be correlated, i.e., the system cannot determine a person's evolution over time. On the other hand, we might overmatch if the window size is chosen too big because of similarities in the addresses or names of people that were actively filing patents in different periods of time.

**Aggregation.** We observe that changing the sketch aggregation window, i.e., the number of hours in a sketch, has minimal impact on the f-measure for this dataset. Specifically, we observe a variation of 0.2% in the f-measure score when varying the sketch size from a day, to a week, a month, and so on up to a year.

**Blocking.** The blocking mechanism that we use for our experiments is the same as used in previous temporal record linkage work. For comparison, we then evaluated single-source processing for our best-case parallel setup with SEQU. Here, we observe a drastic drop in f-measure to 0.09 due to very low precision scores (0.05). Recall that the applied clustering function is a weighted transitive closure variant. As a result, noise such as different people living at similar addresses at different times may distort the output stories. For example Jean-Pierre Gesson lives in Rue Alexandre Dumas which is the same street Jean-Pierre Goedgebuer lives in. This indicates a very interesting property of a (correctly) applied blocking function: In addition to improving performance, it may also help to improve result precision by providing an implicit noise filter, i.e., a smaller chance to wrongly match records or events. Single-source, sequential execution on the other hand may introduce additional noise as shown in this example and observed in these experiments. We also note that artificially splitting a cluster over multiple data sources does not influence whether or not we can link them. However, they are then connected as aligned stories rather than intra-source stories which is an advantage of STORYPIVOT compared to traditional

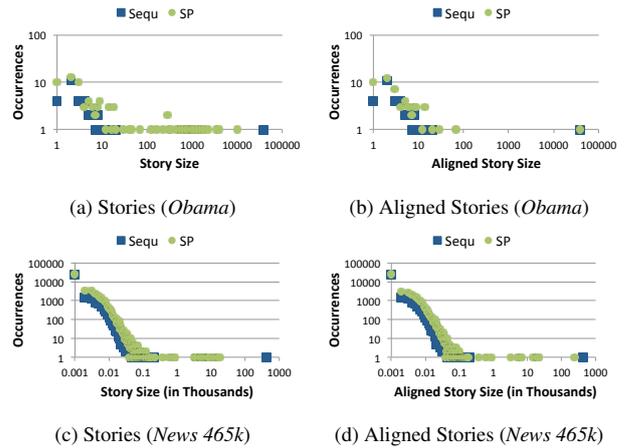

(a) Stories (*Obama*)  (b) Aligned Stories (*Obama*)

(c) Stories (*News 465k*)  (d) Aligned Stories (*News 465k*)

Figure 9: Distribution of (aligned) stories for STORYPIVOT and sequential processing on the *Obama* and *News 465k* datasets.

blocking algorithms: Our system may recover correlations even if the blocking mechanism made errors in the partitioning.

### 6.2.2 Comparison with Sequential Execution

We compare the identified (aligned) stories of STORYPIVOT and the sequential processing model in Figure 9 for the *Obama* and *News 465k* datasets. For the smaller *Obama* dataset, we observe that the distribution of aligned stories produced by STORYPIVOT closely resembles the stories identified by sequentially processing the data. This suggests that given a topic-specific dataset (recall that this dataset was generated by filtering dimension *people* for mentions of token 'Obama'), the parallelization efforts of STORYPIVOT do not diminish the output clustering. Looking at the more general *News 465k* dataset, we observe a difference between the found aligned stories in STORYPIVOT compared to the story distribution of the sequential system. The reason for that is analogous to the observations that we made for the *Patents* dataset: Noise in the dataset often causes erroneous story merging amplified by transitive closure as clustering method. That is, because clusters are linked with noisy edges, they get merged even though they would otherwise be separated and since there are more options to link a sketch with in SEQU, the level of noise increases. Separating the computation of stories into two phases thus reduces the risk of noise amplification. Because of that, we observe of that STORYPIVOT identifies 244,670 snippets that mostly focus on American politics while the largest story identified with SEQU contains 408,404 snippets where some of them are categorized differently in STORYPIVOT (for example they are then classified as American sports events).

### 6.2.3 Comparison with EventRegistry

For the comparison with the window online clustering algorithm of EventRegistry [4], we manually compared the clustering results of the *Obama* dataset with the clusters proposed by EventRegistry. EventRegistry aims to identify real-world events, i.e., mentions of the same event occurring across multiple data sources. In contrast, STORYPIVOT aims to identify the continuous evolution of events over time and capture them through the concept of stories. We therefore expect that it includes the same clusters as EventRegistry in the same stories, thus encompassing them. In addition, it should construct clusters across time and align stories across data sources. In our experiments, we observe that STORYPIVOT identifies 91.52% of the event clusters that have also been automatically computed in EventRegistry. As we do not know EventRegistry's internal weighting scheme, we use in this experiment a uniformly weighted similarity computation across dimensions *people*, *organizations*, and *locations*. A comparison of the story clusters with EventRegistry clusters (i.e., their notion of an 'event'), is shown manually through two examples next.

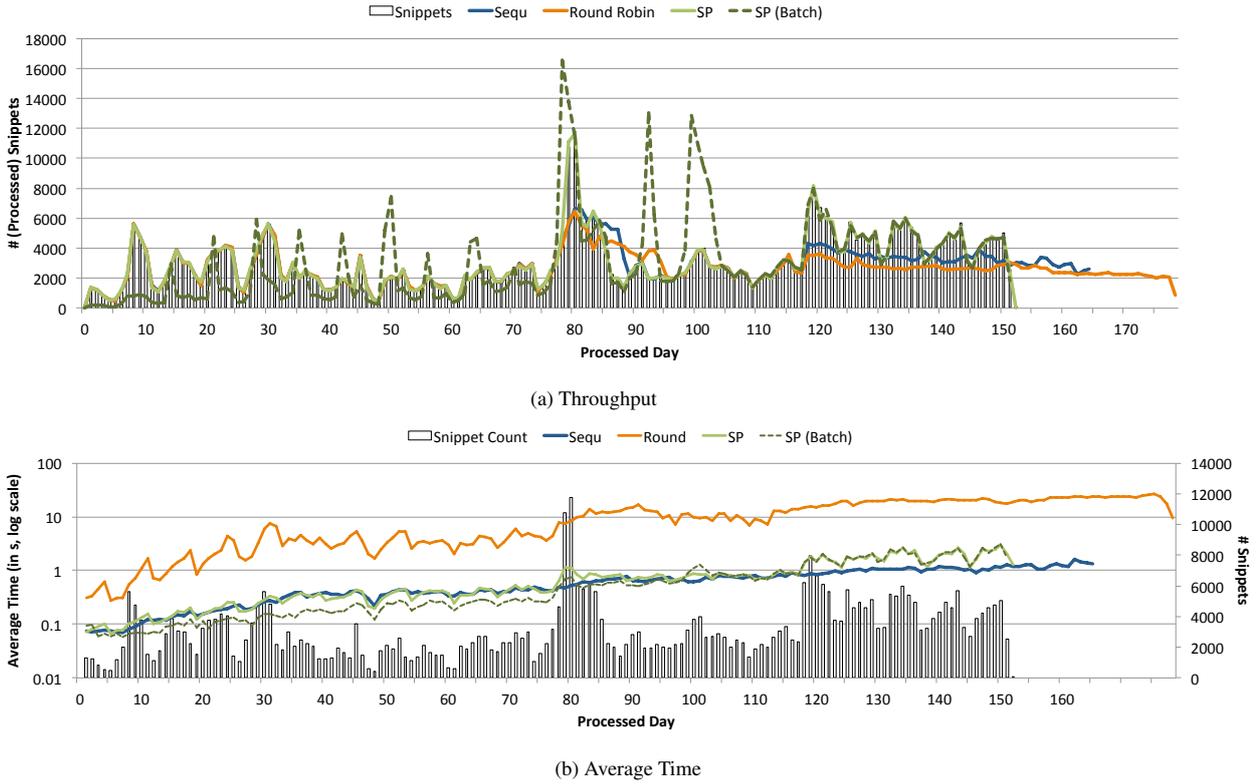

(a) Throughput

(b) Average Time

Figure 10: Throughput and average time for end-to-end snippet processing with varying processing techniques on the *News 465k* dataset.

**Greek Financial Crisis.** After a new Greek government came into power in January 2015, Greece started to negotiate their debt payments in early February. These negotiations were a continuous process that developed over several weeks. In the *Obama* dataset, we observe several extracted snippets in that timeframe that are associated with entities 'Greece', or 'Tsipras' (the Greek prime minister), discussing these negotiations. For example on February 8th, the Reuters news agency published an article (captured through snippet $a_1$) that describes how USA is urging the euro zone to compromise with the Greek government. Two days later the Washington Post reports about the new restrictions proposed by the German chancellor ($a_2$). As these two events are close in time, EventRegistry and STORYPIVOT assign them to the same event (resp. aligned cluster). However, the second news article published in the Washington Post was in fact an update of an article that had been published a week before on February 3rd. Comparing its assigned events according to EventRegistry, we observe that it identified different events for $a_1$ and $a_2$. STORYPIVOT on the other hand discovered that the dimensions were a close match and thus assigned the corresponding snippets to the same story. This example shows the fundamental differences of the conceptual elements in either system, the differentiation between events defined for a specific point in time and stories which span longer time intervals.

**US Immigration Laws.** Similar to the previous example, we observe that EventRegistry correctly matches events across data sources within short time windows. For example, the dataset contains two snippets published in USA Today and the New Zealand Herald about the proposed US immigration laws in mid November 2014 ($b_1$ and $b_2$). On November 21st, the Herald then followed up with an interest piece on what these laws would entail ($b_3$) while USA Today published an article about Obama's point of view of the immigration laws on November 23rd ($b_4$). As $b_3$ and $b_4$ are different in their intent, EventRegistry does not assign them to the same event and fails to draw the connection to $b_1$ for $b_4$ resp. $b_2$ for $b_3$. STORYPIVOT on the other hand not only assigns $b_1$ and $b_4$ ($b_2$ and $b_3$) to the same story but also notes that both stories are related with a similarity across dimensions of 0.357 which is significantly higher than the story alignment threshold with $\alpha_c = 0.1$.

### 6.3 Performance

In this part of our work, we compare the system performance of STORYPIVOT as well as the impact of performance on integration results with competing event processing techniques. First, we show the throughput and average end-to-end integration time of these systems (*End-to-End Experiment*) and discuss performance characteristics for selected data sources (*Data Source Drill-Down*). Afterwards, we vary different parameters to determine the scalability of STORYPIVOT (*Parameter Experiments*).

#### 6.3.1 End-to-End Experiment

In this experiment, we varied the event processing model and recorded throughput and end-to-end integration time per snippet. Figure 10 shows both performance metrics per processed day of the

*News 465k* dataset. In addition to the streaming execution of STO-RYPIVOT, we also record the behavior of a batch implementation where two (randomly selected) data sources are added to the system every seven days (i.e., every seven hours in the actual execution) until the system has integrated all data sources.

**Throughput.** All processing models are able to process snippets in a timely manner for the first part of the dataset, i.e., they are able to process incoming snippets within the same time interval they were issued in. For the batch execution of STORYPIVOT, we observe periodic spikes in the throughput of the system every seven days which reflect the catch-up phase of the system. Once the batch version has integrated all data sources, it has the same performance characteristics as the streaming version of STORYPIVOT. When the number of snippets to process becomes large, we see processing difficulties for both ROUND and SEQU. Specifically, comparing STORYPIVOT with SEQU, we observe a higher throughput of up to 2x for the *News 465k* and up to 4.2x if the same experiment is executed on the *News 1.2m* dataset. The reason is obviously that the system simply cannot process the sheer number of snippets sequentially and thus fails to integrate them in time. For ROUND, the decrease in system performance is due to parallel accesses to the same portions of the indexing structure. Due to consistency, the same object cannot be used by multiple threads in write operations which becomes a problem if multiple threads want to modify (parts of) the same story.

**Execution Time.** While batch executing STORYPIVOT is the fastest approach initially due to a smaller number of integrated sources, it maintains the same performance as SP after day 105 when no data sources are added to the system any longer. Furthermore, the spikes observed in the throughput measurement are not visible in the execution time which suggests that the integration of new data sources does not interrupt the running streaming integration system. However, we observe a general increase in the processing time per day with a large number of snippets for either SP approach. The reason is again concurrent access to the same objects. As the accessed sketches within the same time interval are constantly modified, threads are blocked when accessing them. This is also the reason for the increased execution time of ROUND. In general, we observe that sequential execution is faster on average because it leverages cache consistency when processing related snippets.

### 6.3.2 Parameter Experiments

We next evaluate how different parameters modify the execution time as well as the distribution of stories in STORYPIVOT. Specifically, we look at two different parameters, the comparison interval for sketches, i.e., how far back in time we go to find sketches that might belong to the same story, and the sketch size, i.e., the time frame in which information is aggregated.

**Varying the Comparison Interval.** Figure 11 varies the comparison interval $m'$ from the default $m'$=15 to 7 and 30 days. We observe that with respect to the default execution, decreasing the interval on average improves the end-to-end processing time while increasing the interval worsens performance. Obviously, this is the expected behavior and shown in Figure 11a for the most contested part of the dataset, processing days 100-150. However, note that the difference between $m'$=15 and $m'$=30 (as well as between $m'$=7 and $m'$=15) is sublinear which suggests that intra-source parallelization is efficient and adds only a small overhead. There exists a difference because we limit the number of concurrent threads in our system to avoid running too many threads in parallel and thus incurring too much thread interference. Next to execution time, we observe a smaller number of overall stories with an increased $m'$ (see Figure 11b). In contrast, decreasing $m'$ to 7 increases the number of stories, i.e., pieces of stories are kept apart. This is intuitive as a sketch matching across more time windows obviously increases the chance that two sketches are found similar enough to merge the corresponding stories. We further varied $m'$ to 50 and 100 to observe the impact on the story assignment of events. As expected, we see that the average number of stories per cluster increases, however, we also observe a flattening effect. Specifically, we see that the average number of events per story increases by 9.1% between $m'$=7 and $m'$=15 and only by 5.4% between $m'$=50 and $m'$=100. This suggests that for this dataset, we are able to connect large stories through relatively small time windows.

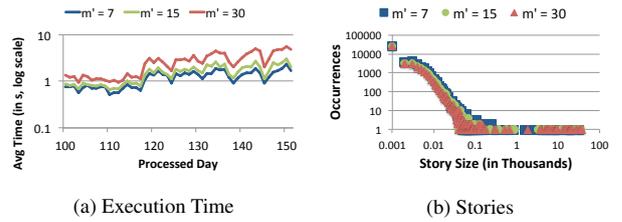

(a) Execution Time    (b) Stories

Figure 11: Comparison of window sizes (*News 465k* dataset).

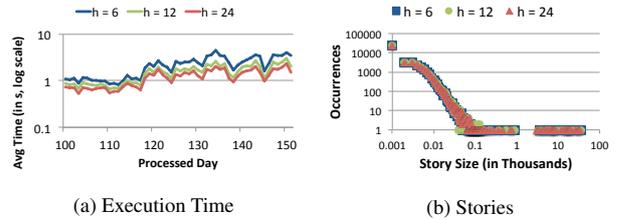

(a) Execution Time    (b) Stories

Figure 12: Comparison of sketch sizes (*News 465k* dataset).

**Varying the Sketch Size.** The observations made for the comparison interval are similar to those made for varying the sketch size when it comes to execution time. Here the default sketch time frame $h$ is twelve hours. In our experiments, we compare it to the execution of $h = 6$ and $h = 24$. We observe that a larger sketch size decreases the execution time as shown in Figure 12a for days 100 to 150 of the *News 465k* dataset. The reason is that fewer comparison objects are injected into the system as more sketches are merged together. As a result, intra- and inter-source look-ups are more efficient. Furthermore, we observe that the distribution of stories does not change when the sketch size is varied as shown in Figure 12b. This indicates that for this specific dataset and parameter, there is no negative trade-off between performance and quality.

### 6.3.3 Scaling StoryPivot

To test the scalability of STORYPIVOT we compare the results of integrating the *News 465k* dataset with those obtained from integrating the *News 1.2m* dataset. In Figure 13 we show the execution time for both datasets comparing our system with a sequential execution. We make three observations. First, the average execution time of SEQU is approximately constant over time if the throughput is maximized (2.02 seconds per snippet which allows the system to process 1782.18 snippets per virtual day). As a result of that, executing the *News 1.2m* dataset takes 444 virtual days while it took 165 virtual days to execute the *News 465k* dataset. Second, STORYPIVOT is not able to process *News 1.2m* within the original 150 virtual days either but requires 206 virtual days less for processing the dataset than SEQU. Around virtual days 120-150, we see an increase in STORYPIVOT's processing time because this is the point in time were multiple data sources produce a high volume of snippets. During that timeframe, we observe that the integration of a snippet requires 15.61 seconds. For comparison, the average execution time per snippet in the *News 465k* dataset is 1.91 seconds for this timeframe. However, we observe that during that time, the throughput of STORYPIVOT is 6986.94 snippets. That is approximately 4 times as much as the sequential algorithm can process within a virtual day (i.e., one execution hour) which is a non-linear improvement over SEQU due to the additional concurrency protocols as well as cache inconsistencies which SP triggers. Third, with nearly 7000 snippets per hour even under high system load, SP can process approximately 150,000 snippets per day. For *News 1.2m*, we observe at most 31,455 snippets in one day which is easily manageable by SP if processed at real-time streaming pace. Sequential processing on the other hand leads to approximately 42,000 snippets processed per day which leaves little space for adding more data sources. Finally, we observe that the execution time increases for highly contested parts of the execution. This behavior is due to the same issue we described above, i.e., data structure consistency as guaranteed by locking mechanisms.

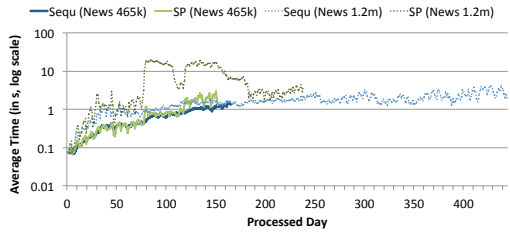

Figure 13: Execution time of STORYPIVOT on both *News* datasets.

## 7. DISCUSSION & FUTURE WORK

We now discuss how the design decisions made when creating STORYPIVOT can be generalized for a wider range of data integration scenarios. For that purpose, we first summarize the contributions that we make in our work. Afterwards, we will analyze how applicable these design features are for other data integration tasks.

**Source-Centric Design Model.** Our system leverages the observation that processing data within a data source first before integrating it across data sources positively impacts the consistency of stories within the data source while it enables us at the same time to improve end-to-end integration time.

**Scalable Integration Mechanisms.** We showed how integration tasks can be parallelized within and across data sources to improve end-to-end integration time.

**Elastic System Design.** We discussed and showed in our experimental evaluation how STORYPIVOT adjusts to new data sources online and without interrupting the running system.

These three design features can be extended in a straightforward manner to other online integration systems that are centered around linking records. Take as an example a data cleaning system which uses syntactic cleaning mechanisms. Commonly, data extracted from different source systems has different and source-dependent quality characteristics. Imagine for example two sources capturing information on employees. One of the systems uses a standardization template that shortens employee addresses (i.e., Mission Street becomes Mission St) while the other system uses the full street name.

This is a traditional example for an ETL process where integrating the data from both data sources leads to an alignment of the unified data set based on some preset cleaning rules. Implementing the design features that we have discussed before for STORYPIVOT is intuitive for such a system: First of all, applying a cleaning rule to a dataset can be easily pushed to each data source without decreasing output quality. Second, this task is parallelizable as each cleaning process can be executed in parallel. Third, alignment across data sources becomes easier as the number of items to compare and consolidate is smaller after in-source standardization.

Data cleaning is not the only integration task that can leverage a source-centric design model but it applies to the more general group of systems that use computationally independent integration mechanisms. An example of interdependent computation is graph processing as seen in Pregel-like systems [30]. In contrast, traditional data integration systems apply rules, pair-wise similarity computation, or metrics on data items where the item's computation is independent of the computational efforts the same system makes on a different data item. Using mechanisms such as blocking that partition the search space (analogous to the time windows within STORYPIVOT), we can leverage parallelization within a data source. Additionally, if there exist multiple data sources, inter-source parallelism can be applied. For this type of integration systems, the design characteristics that we have exemplified with STORYPIVOT can be implemented with minimal overhead. Furthermore, there are a variety of algorithmic and methodological research problems that can be examined in systems such as STORYPIVOT. We therefore discuss two example research directions next, temporal clustering and story visualization, which are useful when constructing this kind of multi-source data integration system.

**Temporal Clustering.** Temporal record linkage is a topic that has been studied in recent years in the context of entity resolution [13, 28]. Here, the idea is that data may change over time which is the same notion that we associate with stories. One area of research that follows up on our work is how to do the (incremental) clustering process that we propose for connecting sketches more accurately. So far, most of the similarity-based computations are based on well-established string metrics. However, as we have pointed out earlier, dimensions may vary in their content (per snippet). Thus, an interesting direction for future research would to learn how to correlate snippets across dimensions when trying to maximize result quality for various temporal, incremental clustering techniques.

**Story Visualization.** STORYPIVOT uses a customized inverted index to map the content of a snippet to the current set of sketches. One of the positive side effects of this design decision is that user queries can use the same access patterns when requesting stories from the system. Given that an aligned cluster can span multiple data sources and large time windows, an obvious question to address in this context is how clusters are visualized once the relevant sketches have been retrieved. We have shown in the experimental evaluation that clusters may contain hundreds of thousands of snippets. Thus, summarization as well as prioritization are important research aspects if STORYPIVOT would be commercialized.

Finally, STORYPIVOT as of now is a single-server application. Thus the number of data sources that can be processed in parallel is limited by the server's capacity. To avoid overloading a server, we next show how our system can be extended to integrate events in a distributed setup. We shortly outline the distributed system next.

**Distributing STORYPIVOT.** Horizontal partitioning of data sources across multiple servers is possible in STORYPIVOT by design. That is, to achieve load balancing and optimal resource utilization per server, the distributed version of STORYPIVOT, DISTRIBUTED STORYPIVOT, assigns different data sources to different inter-connected servers. Multiple data sources may also be assigned to the same server as the number of data sources is expected to exceed the number of servers in the system. For example the *News* dataset used for our experiments in Section 6 contains 24,314 data sources thus a balanced distribution of them is imperative. Figure 14 visualizes DISTRIBUTED STORYPIVOT. Here, the system consists of multiple servers that are used as storage and computation units. Each of the servers is assigned a (subset of) the data sources that the system manages. DISTRIBUTED STORYPIVOT then handles the following three use cases. First, the integration of a snippet into the system; second, the addition of a data source; and third, user requests. The integration of snippets is handled analogous to the single-server version of DISTRIBUTED STORYPIVOT. The only addition to the distributed system is a central management unit that is able to route snippets to their corresponding data source (here $r_7^1$ to $s_1$) and that enables load balancing across servers. The load balancing component decides which server the new data source will be assigned to and will route future requests accordingly. Integration then leverages the same mechanisms described in Section 5. Specifically, intra-source computation is solely executed on the server that $s_1$ resides on after which inter-source communication is initiated for data sources on the same server as well as across the network. Finally, user requests now do not access a single server but need to be propagated which means that the results of these distributed requests need to be aggregated. We envisage that this aggregation can be either done by the central management unit or on one of the system servers that currently has low load. The key take-away here is that extending STORYPIVOT to a distributed system is relatively straightforward because of its source-centric design characteristics. The description of the distributed system given here is a starting point for such a system extension which will come with a set of new challenges such as load balancing, message passing, and more.

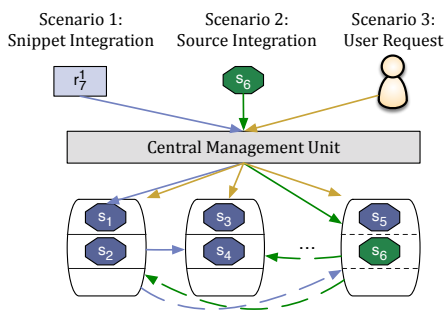

Figure 14: STORYPIVOT as a distributed system.

## 8. RELATED WORK

In this section, we discuss related work in three different areas of research, traditional data integration systems, event processing systems, and streaming systems.

**Data Integration Systems.** There has been a trend in recent years to move away from batch processing in data integration systems towards incremental and iteratively updated integration solutions. In that context, the idea that data changes dynamically and thus the integrated results have to change dynamically has been explored in a variety of research work [5, 22, 27]. This approach differs from traditional data integration systems which assumed static datasets [16, 18] that were integrated over a variety of data sources with a variety of integration algorithms such as entity resolution, deduplication, data cleaning, and others. To model these dynamic changes, algorithms have been furthermore adapted to take source-specific characteristics such as reliability into consideration when making an integration decision [17, 33, 35, 38]. Our work leverages all of these principles and uses them to build an adaptive online integration system. Specifically, we use source-specific characteristics to enable efficient data integration with algorithms that use the same basic principles as incremental data integration algorithms. In contrast to static datasets or datasets that are occasionally updated, we expand the notion of data integration to an online streaming system.

**Event Processing Systems.** Repositories for real-world events such as GDELT [11] or EventRegistry [26] store structured information on real-world events extracted from news articles in a variety of online newspapers. The goal of these systems varies from tracking political unrest to showing current and trending events. A popular use case for such systems is political science which tries to predict political unrest, crisis, and conflicts [25, 36, 39]. Furthermore, financial sciences have shown an increasing amount of interest in event processing systems because of their ability to help understanding and drawing connections between real-world events [24]. Detecting events that have not been previously identified is another line of research that is relevant to our work. Here, research aims to characterize events [34] or to learn about existing events [3] to discover when new events occur. In contrast, the goal of STORYPIVOT is to show the evolution of stories over time to help users understand news articles and to let expert users explore stories within and across data sources [1]. Furthermore, we discuss in our work how we can design event processing systems from scratch to be able to handle complex integration techniques efficiently and how these techniques are generally applicable for a variety of data integration systems and not only limited to event processing.

**Streaming Systems.** Streaming systems [8, 32] have been an increasingly important area of research over the last years because of the sheer amount of data that systems have to process in near real-time. Here, related work has focused on applying traditional database methodology such as joins and top-k processing on data streams [20, 31]. In contrast, our work aims to solve the problem of continuously integrating and maintaining a large amount of data where the data may come in through data streams but needs to be integrated analogous to traditional data integration systems. At the same time, we allow the system to maintain multiple channels of communication similar to the execution of windowed joins on multiple streams simultaneously such as discussed in [20]. Aggregation in the context of streaming systems has been explored for example in [7, 15, 37]. However, these techniques are most commonly used for estimating aggregation objects such as counts or sums.

## 9. CONCLUSION

In this paper, we presented and discussed the design principles behind STORYPIVOT, an integration system that links events over time and across data sources. We explained the quality and performance characteristics of our system and provided implementation details that describe how quality techniques such as cluster identification or performance techniques such as parallelization are realized within STORYPIVOT. The techniques presented in our work are not only applicable for our system but for a wider range of systems that dynamically integrate data. To the best of our knowledge, this paper is the first that discusses how to design such a linkage system for an online environment where data is modified continuously.